\documentclass[12pt]{iopart}
\usepackage{epsf}
\def\simlt{\lower.5ex\hbox{$\; \buildrel < \over \sim \;$}}

\def\mnras{{\it Month.\ Not.\ Roy.\ Astron.\ Soc.\ }}

\def\prd{{\it Phys.\ Rev.\ D}}

\def\be{\begin{eqnarray}}
\def\ee{\end{eqnarray}}
\def\bi{\bibitem}

\def\lrar{\leftrightarrow}

\begin{document}

\title[Neutrino statistics and BBN]
{Neutrino statistics and big bang nucleosynthesis}

\author{A.D. Dolgov$^{1,2}$, S.H. Hansen$^3$, A.Yu. Smirnov$^{4,5}$}

\address{$^1$ Istituto Nazionale di Fisica Nucleare,
Ferrara 40100, Italy \\
$^2$ Institute of Theoretical and Experimental Physics,
117218, Moscow, Russia}

\address{$^3$ Institute for Theoretical Physics, Univ. of Zurich,
Winterthurerstrasse 190, CH-8057 Zurich,  Switzerland}

\address{$^4$ The Abdus Salam International Centre for Theoretical Physics,
I-34100 Trieste, Italy \\
$^5$  Institute for Nuclear Research, Russian Academy of
Sciences, Moscow, Russia}

\begin{abstract}
Neutrinos may possibly violate the spin-statistics theorem, and hence
obey Bose statistics or mixed statistics despite having spin half.
We find the generalized equilibrium distribution function of neutrinos
which depends on a single {\it fermi-bose} parameter, $\kappa$,  and 
interpolates continuously between the bosonic and fermionic distributions
when $\kappa$ changes from -1 to +1. 
We consider modification of the Big Bang Nucleosynthesis (BBN) in the 
presence of bosonic or  partly bosonic neutrinos.  
For pure  bosonic neutrinos the abundances change 
(in comparison with the usual Fermi-Dirac case) by $-3.2\%$ for  
$^{4}$He  (which is  equivalent to a decrease of the effective number of 
neutrinos by $\Delta N_{\nu} \approx - 0.6$), 
$+2.6\%$ for $^{2}$H and  $-7\%$ for $^{7}$Li. 
These changes provide a better fit to the BBN data. 
Future BBN studies  will be able to constrain the fermi-bose parameter
to $\kappa >  0.5$, if no deviation from fermionic nature of neutrinos is
found. We also evaluate the sensitivity of  future  
CMB and LSS observations to  the fermi-bose parameter.

\end{abstract}


%\submitto{\JCAP}

\eads{
\mailto{dolgov@fe.infn.it},
\mailto{hansen@physik.unizh.ch},
\mailto{smirnov@ictp.trieste.it}}

\maketitle

\section{Introduction\label{s-intro}}

Since neutrinos have spin one half they are believed to obey Fermi
statistics. A serious argument in favor of this belief is an absence
of a consistent quantum field theory of half-integer spin particles
with any other statistics than the Fermi one. 
For electrons and nucleons this issue has been discussed
earlier in ref.~\cite{kuzminetal1,kuzminetal2,kuzminetal3,kuzminetal4}, 
where it was shown that a possible violation of
statistics for these particles is strongly restricted by experiment.
On the other hand,
direct experimental checks of Fermi statistics for neutrinos are
absent, except for a study of the effects of purely bosonic neutrinos
on Big Bang Nucleosynthesis (BBN) performed 10 years ago~\cite{gri}.
Recently the idea that neutrinos may possess Bose or mixed statistics
reappeared in ref.~\cite{ad-as} where phenomenological analysis of
testable effects has been presented. A violation of the
spin-statistics relation for neutrinos would lead to a number of
observable effects in cosmology and astrophysics. In particular,
bosonic neutrinos might compose all or a part of the cold cosmological
dark matter (through bosonic condensate of neutrinos) and
simultaneously provide some hot dark matter~\cite{ad-as}. A change of
neutrino statistics, would have an impact on the evolution of
supernovae and on the spectra of supernova neutrinos.  The presence of
a cosmological neutrino condensate would enhance contributions of the
Z-bursts to the flux of the ultra high energy (UHE) cosmic rays and
lead to substantial refraction effects for neutrinos from remote
sources~\cite{ad-as}.

As shown in~\cite{gri} the change of the neutrino statistics from
Fermi-Dirac (F-D) to Bose-Einstein (B-E) leads to a decrease of the
$^{4}$He abundance produced during BBN by about 4{\%}.  The results of
ref.~\cite{ad-as} are in qualitative agreement with ref.~\cite{gri},
however, according to the estimates of ref.~\cite{ad-as} the change of
$^{4}$He is somewhat weaker.

Since the double beta decay excludes the possibility of pure bosonic 
neutrino~\cite{2-beta},
but still allows mixed statistics, we will study here 
an influence of the partially bosonic neutrinos on BBN.
We introduce the {\it fermi-bose} parameter, $\kappa$,  
which describes the continuous transition from Fermi to Bose
distributions in thermal equilibrium as
$\kappa$  changes from $-1$ to $+1$, the boundary cases corresponding to 
purely fermionic and bosonic states with any value of $\kappa$ 
in between allowed. In particular, 
$\kappa = 0$ corresponds to Boltzmann statistics. 
We consider possible constraints on the parameter  
$\kappa$ from present and future BBN and cosmic microwave background
(CMB) plus large scale structure (LSS) data.

\section{The generalized distribution function}
%%%%%%%%%%%%%%%%%%%%%%%%%%%%%%%%%%%%%%%%%%%%%%%%%%%%%%%%%

The form of the kinetic equation for particles with mixed statistics is
not immediately evident. The statistics dependent factor $F[f]$ under
the collision integral  e.g. for the reaction $ 1+2 \lrar 3+4$
in the standard case of pure Bose or Fermi statistics
has the form:
\begin{eqnarray}
F &=& f_1(p_1) f_2(p_2) \left( 1\pm f_3 (p_3) \right) 
\left( 1\pm f_4 (p_4) \right) \nonumber \\
&& - f_3(p_3) f_4(p_4) \left( 1\pm f_1 (p_1) \right) 
\left( 1\pm f_2 (p_2) \right) \, ,
\label{F}
\end{eqnarray}
where the $\pm$ signs  correspond to bosons or fermions. 

Derivation of the kinetic equation 
in general 
depends on the operator of particle number density 
and the normalization of states with non-zero
number of identical particles. 
The quantum creation-annihilation
operators of neutrinos obeying mixed statistics
seems natural to write in the form:
\be
a_k &=&a_k^f \cos \gamma  + a_k^b \sin \gamma, \nonumber\\
a^+_k &=&a^{f+}_k \cos \gamma  +a^{b+}_k \sin \gamma,
\label{fb}
\ee
where $a_k^f$ and $a_k^b$ are respectively the Fermi and Bose type 
operators annihilating states with momentum $k$. 
However, the  particle number operator in the standard form, $n_k = 
a^+_k a_k$, with  $a_k$ defined in (\ref{fb})  is not
satisfactory if there are several identical particles with the same momenta.
The emerging problems can be easily observed if one considers the matrix
element of the particles number operator, $n_k$, between the multiparticle
states defined in the standard way, $ (a^+_k)^N |0\rangle$.  
This state is not an eigenstate of the operator $n_k$. 
Moreover, it is unclear if there 
exists an operator of the  particle number  with the appropriate 
properties, in particular 
commuting with the free Hamiltonian. It may even be that one has to 
abandon the Hamiltonian approach for description of mixed statistics 
particles. Thus, at this stage we can only make a reasonable guess about
the function $F$. 

One simple possibility which has the correct limiting 
behavior in the case of pure statistics is to suggest that the
neutrino distribution function in the final states of any reaction
enters $F$ as
\be
g_\nu \equiv \cos^2 \delta (1 - f_\nu) + 
\sin^2 \delta  (1 + f_\nu) \, ,
\label{1way}
\ee
where $\delta$ is the statistics
mixing angle and $f_\nu $ is the neutrino distribution 
function, which in equilibrium 
interpolates between the F-D and B-E statistics, so that 
$f_\nu^{(eq)} = f_{FD}$ for $\delta = 0$ 
and $f_\nu^{(eq)} = f_{BE}$ for $\delta = \pi/2$.  

The  natural assumption leading to eq. (\ref{1way}) is that the 
outgoing  neutrino contributes 
partly with a Pauli blocking, $1- f_\nu$, with weight $\cos^2 \delta$, and
partly with Bose enhancement, $1+f_\nu$, with weight $\sin^2\delta$.

The angle $\delta$ introduced above should somehow be connected
to the angle  $\gamma$ in the operator mixture (\ref{fb}). The equality  
$\delta = \gamma$ looks as a reasonable hypothesis. If this is true then
the angle $\gamma$ which parametrizes a violation of neutrino statistics
in double beta decay can be constrained from BBN and vice versa. If however,
the angles $\gamma$ and $\delta$ are different the bounds from BBN and 
$2\beta$-decay are not directly related.

The distribution (\ref{1way}) can be rewritten as 
\be
g_\nu \equiv 1 - \kappa f_\nu(\kappa), 
\label{2way}
\ee
where   
\be
\kappa \equiv \cos^2\delta - \sin^2\delta = \cos 2\delta
\ee
which we call the {\it fermi-bose} parameter. 
Thus, we suggest that the distribution of neutrinos in the final state
enters the factor $F$ in the combination (\ref{2way}).
For example, in the case of elastic scattering 
$\nu_1 + l_1 \lrar \nu_2 + l_2$ the neutrino distribution functions 
with mixed statistics appear in the collision integral as
\be 
F =&& f_\nu (k_1) f_l (p_1) [1-f_l (p_2)] 
\left[1 - \kappa f_\nu(k_2)\right] \nonumber \\ 
&& - f_\nu (k_2) f_l (p_2) [1-f_l (p_1)]
 \left[1 - \kappa f_\nu(k_1)\right] \, .
\label{I-coll}
\ee 
The same factor $(1-\kappa f_\nu)$ appears (instead of $(1-f_\nu)$) 
in any process involving mixed statistics neutrinos.

Considering the processes of neutrino scattering and production 
we have found that the factor $F(f_{\nu})$,  and consequently, the  
collision integral, vanish if 
\be 
f_{\nu} = f^{(eq)}_\nu = \left[ \exp (E/T) + \kappa \right]^{-1} \, ,
\label{f-mixed}
\ee 
while all other particle distributions are given by the standard equilibrium 
Bose or Fermi functions. That means that $f^{(eq)}_\nu$ given by 
eq.~(\ref{f-mixed}) is 
the equilibrium distribution function for the case of mixed 
statistics of neutrinos.  

For $\kappa =  +1$  the function $f^{(eq)}$ turns into the Fermi 
distribution, for $\kappa = -1$
it turns into the Bose 
equilibrium distribution, while for $\kappa  =0$, {\it i.e.}  for equal 
mixture of Fermi  and Bose statistics, it becomes the Boltzmann one.

\begin{figure}[hbt]
\begin{center}
\epsfxsize=13cm
\epsfysize=10cm
\epsffile{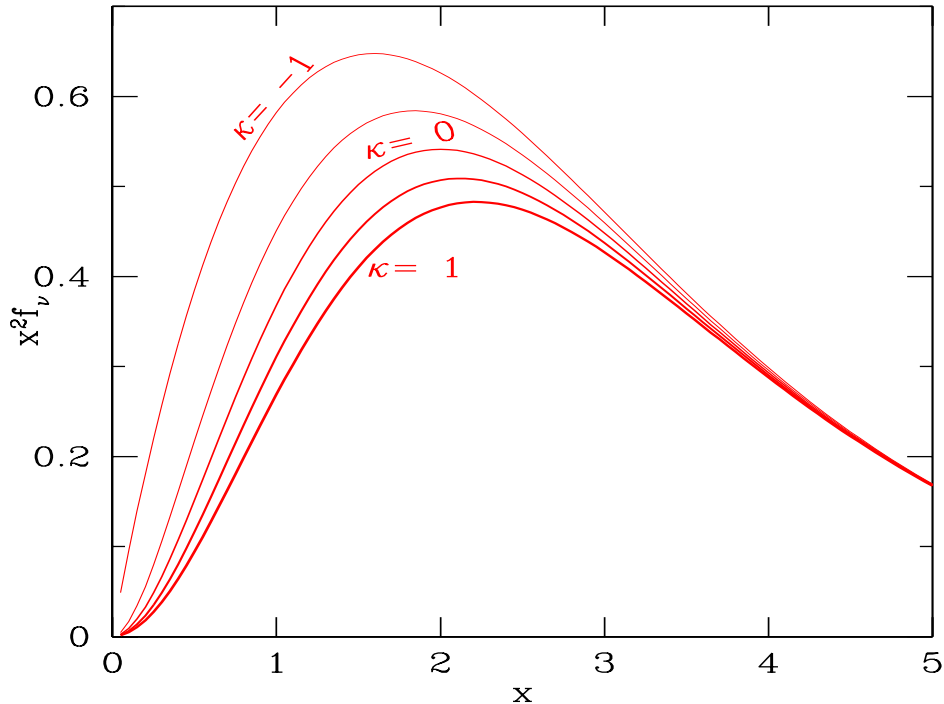}
\caption{The neutrino distribution function for different values 
of $\kappa$ $(-1,-0.5,0,0.5,1)$. Here $x = E/T$.  
}
\label{ffnu}
\end{center}
\end{figure}

In fig. \ref{ffnu}  we present the equilibrium distributions 
(\ref{f-mixed}) for different  values of 
$\kappa$. According to this figure the distribution becomes softer
with an increase of the 
bosonic fraction. The maximum number density shifts 
to smaller $E/T$, and the integrated number density increases.

%%%%%%%%%%%%%%%%%%%%%%%%%%%%%%%%%%%%%%%%%%%%%%%%%%%%%%%%%%%%%%%%%
%%%%%%%%%%%%%%%%%%%%%%%%%%%%%%%%%%%%%%%%%%%%%%%%%%%%%%%%%%%%%%%%%%%%%%%

For a positive $\kappa$ (when the fermionic component dominates) the 
distribution function $f_\nu$ is bounded from above by $f_\nu < 1/\kappa$.

If $\kappa$ is negative, then for a large lepton asymmetry a neutrino  
condensation would be possible. Indeed, for a given $\kappa$, the maximum 
allowed value of the chemical potential is 
\be
\mu^{(max)} = m_\nu - T \ln (-\kappa) \, ,
\label{mumax}
\ee 
this follows from the condition that $f_\nu$ should be non-negative.
In particular, for the purely bosonic case we obtain the usual bound 
$ \mu \leq m_\nu$.
 
If the neutrino charge asymmetry is so large that $\mu^{(max)}$ could
not provide it,  the neutrinos would form a Bose condensate with the
equilibrium distribution function equal to
\be
f^{(eq)} = \frac{1}{ \exp (E -\mu^{(max)} )/T + \kappa}
+ C \delta ({\bf k}) \, .
\label{f-chem}
\ee 
Here ${\bf k}$ is the three momentum of the neutrino, and $C$ is
a constant whose magnitude is determined by the value of the charge
asymmetry of the neutrinos. 

An interesting question to address is how unique is the 
distribution (\ref{f-mixed}). Can one introduce other 
forms of the mixed statistics equilibrium distributions? 
An alternative to (3) could be 
\be
g_\nu \equiv \cos^2 \delta (1 - f_{FD}) +
\sin^2 \delta  (1 + f_{BE}) \, .
\label{3way}
\ee
Such kinetics is equivalent to having two independent neutrino species,
bosonic and fermionic, and, if equilibrium is established, 
it would lead to $24/7$ neutrino species (see next section). 
To avoid equilibrium the bound on $\delta$ in this case 
would be quite restrictive. 

%%%%%%%%%%%%%%%%%%%%%%%%%%%%%%%%%%%%%%%%%%%%%%%%%%%%%%%%%%%%%%%%%
\begin{figure}[hbt]
\begin{center}
\epsfxsize=14cm
\epsfysize=12cm
\epsffile{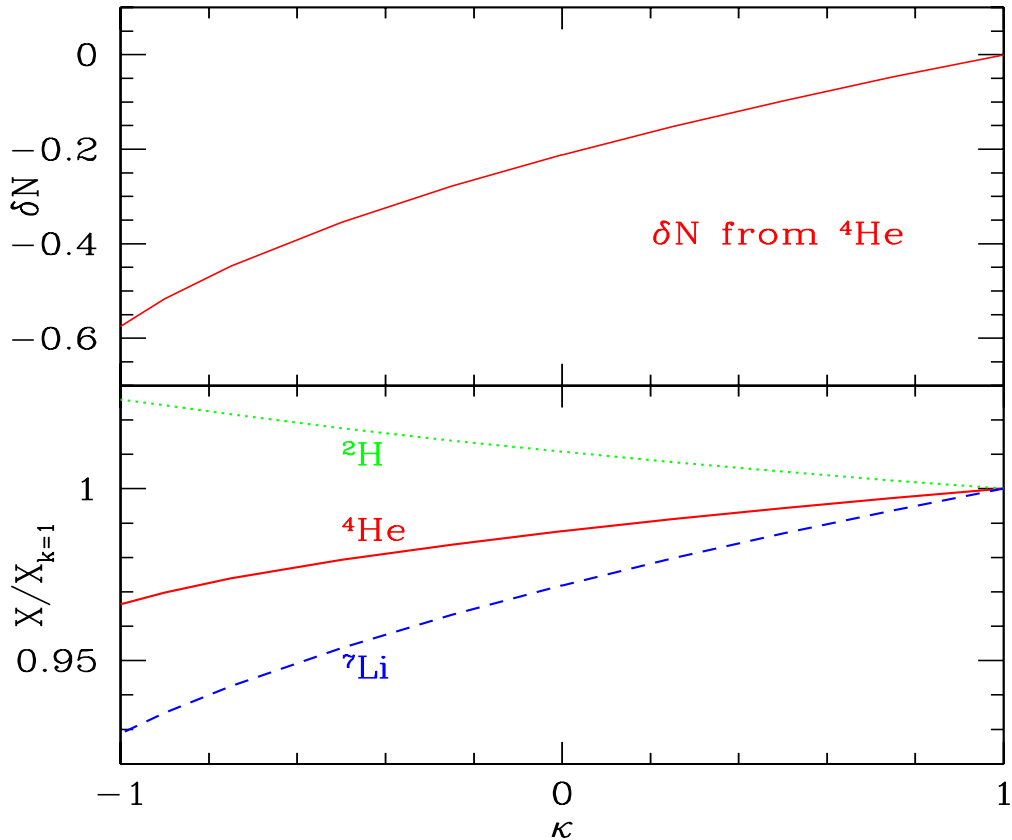}
\caption{Upper panel: the change in the effective number of 
neutrino degrees of freedom found from
the change of the $^4$He abundance as a 
function of the effective fermi-bose parameter $\kappa$.  
Lower panel: the relative change of the
primordial abundances of deuterium, helium-4, and lithium-7, as 
functions of the effective fermi-bose coefficient $\kappa$. 
We take $\eta = n_B/n_\gamma =6.5 \cdot 10^{-10} $.
}
\label{fig:k}
\end{center}
\end{figure}
%%%%%%%%%%%%%%%%%%%%%%%%%%%%%%%%%%%%%%%%%%%%%%%%%%%%%%%%%%%%%%%%%%%%%%%

\section{Effects on big bang nucleosynthesis}
%%%%%%%%%%%%%%%%%%%%%%%%%%%%%%%%%%%%%%%%%%%%%%%

The effect of the change of neutrino statistics on BBN is related to
two phenomena. First, the energy density of  bosons in  equilibrium  
($\kappa = -1$) is 
larger than the energy density of fermions by the factor 8/7. 
If all three neutrinos have B-E statistics, 
their larger energy density would correspond to an increase of the
effective number of neutrino species by $\Delta N_\nu = 3/7$. The
second, dominant, effect is an increase of the rate of the
reactions of neutron-to-proton transformations 
due to the bosonic (anti blocking) factor $(1 - \kappa f_{\nu})$. Because 
of the larger rate the  
freezing temperature of these reactions would be lower and 
consequently the frozen
$n/p$-ratio would be smaller. It  can be mimicked by a 
decrease of $N_\nu$.
The first effect can be approximately described as 
$\Delta N_\nu \approx 0.2 (1-\kappa)$, while the second one is noticeably 
non-linear (exponential).

We performed the calculations of the abundances of light elements  with the 
Kawano nucleosynthesis code~\cite{kawano} which was modified to include 
the effects of mixed  statistics of neutrinos described by the 
distribution  (\ref{f-mixed}).  
This code~\cite{kawano} is accurate
enough for calculations of the relative changes of the abundances, while
for the absolute values of the abundances we use the results of 
a more precise modern code~\cite{serpico}. As a
reference value of the baryon number density we take the WMAP
result~\cite{wmap}, $\eta \equiv n_B/n_\gamma =6.5 \cdot 10^{-10} $.

\begin{figure}[htb]
\begin{center}
\epsfxsize=14cm
\epsfysize=12cm
\epsffile{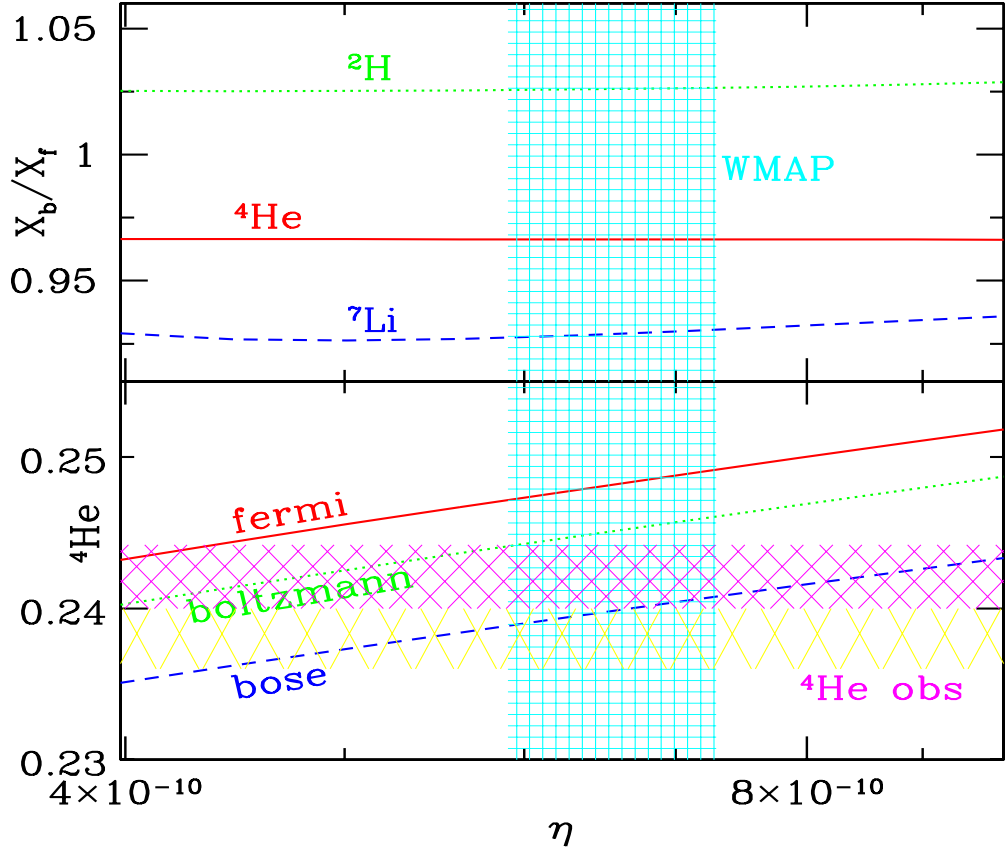}
\end{center}
\caption{Upper panel: the ratios of abundances 
of different elements in the cases of purely
bosonic neutrinos with respect to the standard
fermionic case as functions of
the baryon number density, $\eta$. The vertically hatched (cyan) region 
shows the WMAP $2\sigma$ determination of $\eta$. Lower panel: the
absolute abundance of $^4$He as a function of $\eta$ for the purely
bosonic, Boltzmann, and fermionic neutrino distributions, corresponding
to $\kappa=-1, 0, +1$ respectively. The two skew hatched 
regions show the observation of primordial
helium from ref.~\cite{fieldsolive} (lower, yellow) 
and ref.~\cite{izotovthuan} (upper, magenta), which marginally
overlap at $1\sigma$.}
\label{fig:deut}
\end{figure}

The results of the computations are shown in figs.~\ref{fig:k} and
\ref{fig:deut}. 
In the upper panel of fig. \ref{fig:k} we present the change of the  
effective number of neutrino species, $\delta N_\nu$,
as a function of $\kappa$,
which is equivalent to 
a decrease of the $^4$He primordial  abundance.
If the neutrinos have a purely bosonic distribution ($\kappa=-1$),  
the effect is similar to having $\Delta N_{\nu} \approx -0.57$.  

However, the effect of modified statistics cannot be described by a
simple change in $N_\nu$ if other light elements are included.  In the
lower panel of fig.~1 the relative changes of the abundances of $^2$H,
$^4$He, and $^7$Li with $\kappa$ are shown. As expected the mass
fraction of $^4$He drops down and for pure bosonic neutrinos we get
the relative decrease about \be \frac{^4{\rm He}(\kappa = -1)}{^4{\rm
He}(\kappa = 1)} - 1 = - 3.2\%.  \ee 
This is slightly smaller than the value $-3.7$\% found in \cite{gri}.

The amount of $^2$H goes up with decrease of $\kappa$.  
A higher deuterium abundance can be explained by a slower conversion
rate of deuterium to heavier elements due to fewer neutrons and higher
expansion rate at BBN epoch when $T\approx 0.8 \cdot 10^9$K. In the pure 
bosonic case 
the increase is about 2.6 \%. 
The $^7$Li abundance decreases with $\kappa$, and for $\kappa = -1$
the decrease is 
about 7\%.

%%%%%%%%%%%%%%%%%%%%%%%%%%%%%%%%%%%%%%%%%%%%%%%%%%%%%%%%%%%%%%
%%%%%%%%%%%%%%%%%%%%%%%%%%%%%%%%%%%%%%%%%%%%%%%%%%%%%%%%%%%%%%%%%

Let us confront the absolute values of the 
abundances for partly bosonic neutrinos with 
observational results. We will use the relative changes 
presented in fig.~\ref{fig:k}
and the central values of the abundances 
calculated for usual F-D neutrinos in  ref.~\cite{serpico}: $Y_p =0.2481$,
$X_{^2H}/X_H = 2.44\cdot 10^{-5}$, and $X_{^7 Li}/X_H = 4.9\cdot
10^{-10}$.  Other codes may give even slightly larger values
for the helium abundance (G. Steigman, private communication).
We are here (and in the figures) using the symbol
$X$ for the abundance of $D$ or $Li$ ratios to $H$ by number, not
by mass.
At $\kappa=-1$ we find 
for $^4$He: $Y_p =0.240$, which makes much better agreement 
with the value extracted from 
observations (for a review of the latter see {\it e.g.}~\cite{bbn-obs}).
Different helium observations yield different results, {\it e.g.}, the   
ref.~\cite{fieldsolive} finds $Y=0.238 \pm 0.002$, and ref.~\cite{izotovthuan}
finds  $Y = 0.2421 \pm 0.0021$ ($1\sigma$, only statistical error-bars).
These results are shown in figure 2 as the skew hatched  regions, where 
the upper (magenta) is the results of ref.~\cite{izotovthuan} and the
lower (yellow) is the results of ref.~\cite{fieldsolive}.
Whether the existing helium observations are accurate or slightly
systematically shifted will be tested with future CMB 
observations~\cite{trotta}.  

The amount of $^2$H rises at most to $X_{^2H}/X_H = 2.5\cdot 10^{-5}$, and
the agreement between BBN and WMAP data remains good, 
bearing in mind the observational uncertainties. 
Primordial $^7$Li drops down to  $X_{^7 Li}/X_H = 4.55\cdot 10^{-10}$,
again slightly diminishing the disagreement between theory and 
observations.

We see that at the present time BBN does not exclude even a pure
bosonic nature of all three neutrinos. Furthermore, the agreement
between the value of the baryonic mass density, $\eta$, inferred from
CMBR and the predicted abundances of $^4$He, $^2$H, and $^7$Li becomes
even better. In other words, in the standard BBN model there is an
indication of disagreement between observations of $^4$He and $^2$H -
they correspond to different values of $\eta$ with the observed
abundances of $^4$He indicating a smaller value of $\eta$ than the one
given by CMBR, while $^2$H agrees with CMBR. Motivated by these
results the value of $\Delta N_\nu = -0.7\pm 0.35$ was suggested in
ref.~\cite{steig}.  In the case of predominantly bosonic neutrinos, as
discussed above, the discrepancy between $^2$H, $^4$He, and CMBR
disappears.

When the problem of large systematic uncertainties of primordial helium 
determinations will be resolved  and  the statistical error-bars will 
dominate the error-budget,  one can expect to
measure  $N_\nu$ with an accuracy at the level of 0.1. 
This  would exclude $\kappa < 0.5$,  if an agreement with the standard BBN 
values is found. Otherwise, if the discrepancy between $^4$He and $^2$H 
remains it may be considered as an indication to the mixed statistics of
neutrinos.

Our results change only slightly with variation of the
baryon number density $\eta$, as seen in fig.~2. 
The upper panel shows the ratio of abundances for purely bosonic to
purely fermionic neutrinos. The changes are always of the order a few
percent for the 3 abundances considered.  
The results are in good agreement with ref.~\cite{gri}.
The vertically hatched (cyan) region
shows the $2\sigma$ WMAP result. 
The lower panel shows the absolute
value of the  $^4$He abundance as a function of $\eta$, for purely 
bosonic, Boltzmann,  and
purely fermionic neutrino distribution functions. For other values 
of $\kappa$, the result will be in between those
lines, and can be obtained using curves of fig. \ref{fig:k}. 
The skew hatched 
(yellow) region  shows the range of observed values of 
the helium abundance from refs.~\cite{fieldsolive,izotovthuan},
which marginally overlap at $1\sigma$.\\

It is well known that CMB can be used to constrain the number of
relativistic degrees of freedom at the time of photon decoupling (see
e.g. \cite{hu99}).  
For CMB and  LSS,  in contrast to BBN,  the 
presence of the bosonic 
neutrinos  increases  the number  of 
degrees of freedom from $0$ to 
$3/7 = 0.43$ as $\kappa$ goes from $+1$ to $-1$. 
The present bounds from CMB and LSS data are 
insensitive to such changes and too weak to 
constrain the fermi-bose parameter~(see~\cite{present} for references).
The Planck experiment  is forecast to constrain the
relativistic degrees of freedom to the level 
$\delta N \approx 0.24$~\cite{bowen} at $1\sigma$.
This means that Planck alone will be able to
measure $\kappa$ with a precision of about $\Delta \kappa \approx 1$.
In particular, a pure 
bosonic distribution function  for neutrinos can be  excluded 
at about $2\sigma$ level.    
An ``ambitious'' future experiment (see details in ref.~\cite{cuoco},
and for earlier predictions see~\cite{bashinsky})
will constrain $\Delta N$ to about $0.02$, which corresponds to a
determination of $\kappa$ with precision $\Delta \kappa \approx 0.1$
at $1\sigma$.\\

It is known from cosmological considerations that the mass of
fermionic hot neutrinos are bounded from above by approximately 1 eV,
for recent studies see e.g.~\cite{hep-ph/0504059,hep-ph/0505148}.
This bound is applicable to any hot dark matter particle,
independently of their statistics, which have the same number density
as that of neutrinos.  For particles which have different number
densities, or freeze out at different times, this number changes
somewhat~\cite{hep-ph/0504059}.  Since large scale structure
basically constrains the quantity 
\be \Omega_\nu h^2 = \frac{\sum m_i}{93 eV} 
\, \frac{n_i}{n_{th}} \,
\ee 
where $n_{th}$ is the number density of a thermal fermionic neutrino,
then bosonic neutrinos will have their masses constrained a factor 4/3
weaker than fermionic neutrinos. We thus see that cosmological probes
of neutrino masses remain roughly as strong as always, in comparison
to terrestrial tritium or double beta decay experiments.

In addition to the thermal neutrino component bosonic neutrinos might
condense in the early universe and have much larger number density
than thermal relics, as argued in ref.~\cite{ad-as}. However, the cosmological
upper bound on their mass would remain practically the same because
the latter is valid only for hot thermal relics which suppress
structure formation at small scales.

The higher number density of bosonic neutrinos will imply a marginally
later freeze out, and hence a larger sharing of the entropy from the
annihilating electrons than in the standard
scenario~\cite{hm,dhs,mmpp}. This effect is, however, very small.
More general non-thermal neutrino spectra and the effect on freeze out
is studied in \cite{cuoco}.

Let us comment on a possibility of large neutrino condensate 
in the case of partly bosonic neutrinos. 
Such a  condensate would contribute as cold dark matter in 
the Universe (and show up in the CMB and LSS analysis).  
The condensate is formed when the lepton asymmetry is larger than 
that which could be ensured by
the maximal possible chemical potential. In the case of pure bosonic 
neutrinos the chemical potential is restricted by the neutrino mass and 
is therefore negligible, especially at the BBN epoch. 
In contrast, in the case of partly bosonic neutrinos  
the maximal potential given by (\ref{mumax}),
or $\xi = \mu/T = m_\nu/T - {\rm ln}(-\kappa)$, 
can be large.  So, in the case of partly bosonic neutrinos, the 
formation of the  condensate would imply 
a large chemical potential, which could destroy the excellent agreement
with BBN. 
Due to  mixing between the active neutrinos
the chemical potentials should  be equal  for all 3 neutrino 
species  at the time of BBN~\cite{lunardini,dhpprs}.  
Then using the strong bound on the leptonic asymmetry in the electron 
neutrinos  we find the bound $\kappa < -0.9$: 
For $\kappa > -0.9$ such a chemical potential for the electron
neutrinos will significantly underproduce helium, leading to a
disagreement with observations. For negative chemical potential
$^4 He$ would be strongly overproduced leading to essentially the same 
bound, $\kappa < -0.9$.
 
That is, neutrinos should be  almost purely bosonic to 
produce the condensate and satisfy the BBN bound. 
On the other hand, almost purely bosonic neutrinos are 
excluded (disfavored) by the double beta decay~\cite{2-beta}:
the mixing angle $\theta$ at the level ${\rm sin} \theta \sim 0.8$ is
still allowed, however, the angle $\theta$ is not necessarily equal to
$\delta$ introduced above.
Notice further that the relation of $\kappa$ with the 
fermi-bose parameter relevant for the $\beta\beta$ -decay is not 
clear, as discussed in section~2.

Anyway, an improvement of the BBN bound on $\kappa$ can exclude 
a possibility of the neutrino condensate which might contribute 
substantially to the cold dark matter in the Universe.

\section{Conclusions}
%%%%%%%%%%%%%%%%%%%%%%%%%%%%%%%%%%%%%%%%%%%%%%%%%%%%%%%%

We find the equilibrium distribution function for
partially bosonic neutrinos which depends on  a single {\it
fermi-bose} parameter, $\kappa$. The change of this parameter from
$+1$ to $-1$ corresponds to a continuous transition between Fermi and
Bose distributions.

We have considered the influence of bosonic or partially bosonic
neutrinos on BBN. In the extreme case of completely or predominantly
bosonic neutrinos the primordial abundances change in comparison with 
the usual F-D cases in the following way:   
$^4$He decreases by 3.2\%, $^2$H increases by 2.6\% and 
$^7$Li decreases by 7\%.    
The agreement between theory and observations
becomes noticeably better.

Future determinations of $^4$He will allow to exclude  
values of fermi-bose parameter  $\kappa <  0.5$, if agreement with 
the standard case is found. 
The BBN bounds on
$\kappa$ can be compatible with those obtained from the analysis of
two neutrino double beta decay~\cite{2-beta}.

Future CMB+LSS observations can constrain or
observe this parameter, possibly to the level $\Delta \kappa \approx 0.1$,
potentially providing indications of a
violation of the Pauli exclusion principle.

%Unfortunately
%we cannot rigorously express the parameters describing statistics violation
%in these two sets of data, though it seems reasonable that the Bose-Fermi
%mixing angles $\gamma$ and $\delta$ introduced above in Sec. 2 are the same.

\ack 
SHH thanks the Tomalla foundation for financial
support. 

\label{lastpage}


\section*{References}
\begin{thebibliography}{99}

\bi{kuzminetal1}
Ignatiev A Yu and Kuzmin V A, 1987 {\it Yad. Fiz.} {\bf 46}  786;
%[Sov.~J.~Nucl.~Phys. {\bf 46} (1987) 786]; 
1988 {\it JETP Lett.} {\bf 47} 4

\bi{kuzminetal2}
Okun L B, 1987
%Pis'ma ZhETF, {\bf 46} (1987) 420 
JETP Lett. {\bf 46}  529; 
1988 {\it Yad. Fiz.} {\bf 47} 1192

\bi{kuzminetal3}
Greenberg O W and  Mohapatra R N, 1987 
{\it Phys.~Rev.~Lett.} {\bf 59} 2507;
1989
{\it Phys.~Rev.~Lett.} {\bf 62} 712, {\it Phys.~Rev. D} {\bf 39} 2032

\bi{kuzminetal4}
Govorkov A B, 1989 {\it Phys.~Lett. A} {\bf 137} 7;
Okun L B, 1989 
{\it Uspekhi Fiz. Nauk} {\bf 158} 293;
%[Sov.~Phys.~Usp. {\bf 32} (1989) 543;\\
{\it Comments Nucl.~Part.~Phys.,} {\bf 19} 99


\bi{gri} Cucurull L, Grifols J A, Toldra R, 1996
%SPIN STATISTICS THEOREM, NEUTRINOS, AND BIG BANG NUCLEOSYNTHESIS.
{\it Astropart. Phys.} {\bf 4} 391


\bi{ad-as}
Dolgov A D  and   Smirnov A Yu, 2005 hep-ph/0501066

\bi{2-beta}
Barabash A S,
Dolgov A D, Domin P, Smirnov  A Yu and {\~S}imkovic  F, 2005 in preparation

\bi{kawano}
Kawano L, Fermilab-Pub-92/04-A

\bi{serpico}
Serpico P D {\it et al}, 2004
%, S Esposito, F Iocco, G Mangano, G Miele, O Pisanti,
JCAP {\bf 0412} 010

\bi{wmap}
Spergel  D N {\it et al}, 2003
{\it Astrophys. J. Suppl.} {\bf 148} 175

\bi{bbn-obs}
Eidelman S {\it et al}, 2004 
{\it Phys. Lett.} {\bf B592} 1
%Fields B D  and Sarkar S, 

\bibitem{fieldsolive}
Fields B D and Olive K A, 2005
%``On the Evolution of Helium in Blue Compact Galaxies,''
arXiv:astro-ph/9803297.
%%CITATION = ASTRO-PH 9803297;%%

\bibitem{izotovthuan}
Izotov Y I and Thuan T X, 2004,
%``Systematic effects and a new determination of the primordial abundance of
%4He and dY/dZ from observations of blue compact galaxies,''
{\it Astrophys.\ J.}  {\bf 602}, 200
%[arXiv:astro-ph/0310421].
%%CITATION = ASTRO-PH 0310421;%%

\bi{trotta}
Trotta R and Hansen S~H, 2004
%``Observing the helium abundance with CMB,''
{\it Phys. Rev. D} {\bf 69} 023509
%[arXiv:astro-ph/0306588]
%%CITATION = ASTRO-PH 0306588;%%



\bi{steig}
Steigman G, astro-ph/0501591, 
%BBN and the Primordial Abundances
To appear in the Proceedings of the ESO/Arcetri Workshop on 
"Chemical Abundances and Mixing in Stars in the Milky Way and its 
Satellites", eds., L. Pasquini and S. Randich (Springer-Verlag Series, 
"ESO Astrophysics Symposia").

\bi{hu99}
Hu W, Eisenstein D~J, Tegmark M and White M~J, 1999
%``Observationally Determining the Properties of Dark Matter,''
{\it Phys. Rev. D} {\bf 59} 023512
%[arXiv:astro-ph/9806362]
%%CITATION = ASTRO-PH 9806362;%%

\bi{bowen}
Bowen R {\it et al}, 2002 
%S.~H.~Hansen, A.~Melchiorri, J.~Silk and R.~Trotta,
%``The Impact of an Extra Background of Relativistic Particles on the
%Cosmological Parameters derived from Microwave Background Anisotropies,''
\mnras {\bf  334} 760
%[arXiv:astro-ph/0110636].
%%CITATION = ASTRO-PH 0110636;%%


\bi{present}
Crotty P, Lesgourgues J and Pastor S, 2004
%``Current cosmological bounds on neutrino masses and relativistic relics,''
{\it Phys. Rev. D} {\bf 69} 123007;
%[arXiv:hep-ph/0402049];
%%CITATION = HEP-PH 0402049;%%
2003 \prd {\bf 67} (2003) 123005;
%[arXiv:astro-ph/0302337]
%%CITATION = ASTRO-PH 0302337;%%
Hannestad S, 2004
%``Neutrinos in cosmology,''
{\it New J.\ Phys.}  {\bf 6} 108;
%[arXiv:hep-ph/0404239]
%%CITATION = HEP-PH 0404239;%%
Hansen S~H {\it et al}, 2002
%, G~Mangano, A~Melchiorri, G~Miele and O~Pisanti,
%``Constraining neutrino physics with BBN and CMBR,''
\prd {\bf 65} 023511
%[arXiv:astro-ph/0105385];
%%CITATION = ASTRO-PH 0105385;%%



\bibitem{cuoco}
Cuoco A, Lesgourgues J, Mangano G and Pastor S,  2005,
%``Do observations prove that cosmological neutrinos are thermally
%distributed?,''
arXiv:astro-ph/0502465.
%%CITATION = ASTRO-PH 0502465;%%

\bibitem{bashinsky}
Bashinsky S and Seljak U, 2004,
%``Signatures of relativistic neutrinos in CMB anisotropy and matter
%clustering,''
\prd {\bf 69}, 083002
%[arXiv:astro-ph/0310198].
%%CITATION = ASTRO-PH 0310198;%%


\bibitem{hep-ph/0504059}
Hannestad S, Mirizzi A and Raffelt G, 2005,
%``New cosmological mass limit on thermal relic axions,''
arXiv:hep-ph/0504059.
%%CITATION = HEP-PH 0504059;%%

\bibitem{hep-ph/0505148}
Pastor S, 2005,
%``Massive neutrinos and cosmology,''
arXiv:hep-ph/0505148.
%%CITATION = HEP-PH 0505148;%%

\bibitem{hm}
Hannestad S and Madsen J, 1995,
%``Neutrino decoupling in the early universe,''
Phys.\ Rev.\ D {\bf 52}, 1764.
%[arXiv:astro-ph/9506015].
%%CITATION = ASTRO-PH 9506015;%%


\bibitem{dhs}
Dolgov A.~D., Hansen  S.~H. and Semikoz D.~V., 1997,
%``Non-equilibrium corrections to the spectra of massless neutrinos in the
%early universe,''
Nucl.\ Phys.\ B {\bf 503}, 426;
%[arXiv:hep-ph/9703315].
%%CITATION = HEP-PH 9703315;%%
%``Nonequilibrium corrections to the spectra of massless neutrinos in the
%early universe. (Addendum),''
1999, Nucl.\ Phys.\ B {\bf 543}, 269.
%[arXiv:hep-ph/9805467].
%%CITATION = HEP-PH 9805467;%%

\bibitem{mmpp}
Mangano G, Miele G, Pastor G and Peloso M, 2002,
%``A precision calculation of the effective number of cosmological
%neutrinos,''
Phys.\ Lett.\ B {\bf 534}, 8
%[arXiv:astro-ph/0111408].
%%CITATION = ASTRO-PH 0111408;%%


\bibitem{lunardini}
Lunardini C and Smirnov A Ya, 2001,
%``High-energy neutrino conversion and the lepton asymmetry in the
%universe,''
{\it Phys.\ Rev.\ D} {\bf 64}, 073006
%[arXiv:hep-ph/0012056].
%%CITATION = HEP-PH 0012056;%%

\bibitem{dhpprs}
Dolgov A D et al., 2002,
%``Cosmological bounds on neutrino degeneracy improved by flavor
%oscillations,''
Nucl.\ Phys.\ B {\bf 632}, 363
%[arXiv:hep-ph/0201287].
%%CITATION = HEP-PH 0201287;%%


\end{thebibliography}
\end{document}